\documentclass[aps,letterpaper,twocolumn,nofootinbib,floatfix]{revtex4-1}
\usepackage{natbib}
\usepackage{cancel}
\usepackage{amsthm}         
\usepackage{amsmath} 	   
\usepackage{amssymb}       
\usepackage{amsfonts}
\usepackage{graphicx} 	    
\usepackage{verbatim}
\usepackage{color,colortbl}
\usepackage{float}

\begin{document}

\title{A general mechanism for producing scale-invariant perturbations and small non-Gaussianity in ekpyrotic models} 

\author{Anna Ijjas}
\email{aijjas@princeton.edu}
\affiliation{Max-Planck-Institute for Gravitational Physics (Albert-Einstein-Institute), 14476 Potsdam, Germany} \affiliation{Rutgers University, New Brunswick, NJ 08901, USA}

\author{Jean-Luc Lehners}
%\email{}
\affiliation{Max-Planck-Institute for Gravitational Physics (Albert-Einstein-Institute), 14476 Potsdam, Germany}

\author{Paul J. Steinhardt}
%\email{steinh@princeton.edu}
\affiliation{Department of Physics and Princeton Center for Theoretical Science, Princeton University, Princeton, NJ 08544, USA}

\date{\today}

\begin{abstract}
We explore a new type of entropic mechanism for generating density perturbations in a contracting phase in which there are two scalar fields, but only one has a steep negative potential.  This first field dominates the energy density and is the source of the ekpyrotic equation of state.  The second field has a negligible potential, but its kinetic energy density is coupled to the first field with a non-linear sigma-model type interaction.  
We show that for any ekpyrotic equation of state it is possible to choose the potential and the kinetic coupling such that exactly scale-invariant (or nearly scale-invariant) entropy perturbations are produced. The corresponding background solutions are stable, and the bispectrum of the entropy perturbations vanishes as no non-Gaussianity is produced during the ekpyrotic phase. Hence, the only contribution to non-Gaussianity comes from the non-linearity of the conversion process during which entropic perturbations are turned into adiabatic ones, resulting in a local non-Gaussianity parameter $f_{\textsc{nl}} \sim 5$.  
\end{abstract}

\maketitle

\section{Introduction}

Recent {\it Planck} satellite measurements \cite{Ade:2013lta,Ade:2013rta,Ade:2013ydc}, together with earlier 
observations from WMAP, ACT, SPT, and other experiments \cite{Sievers:2013ica}, showed with high precision that the spectrum of primordial density fluctuations is nearly scale-invariant, Gaussian, and adiabatic. The currently best known mechanisms for the generation of cosmological perturbations are inflation \cite{Guth:1980zm,Linde:1981mu,Albrecht:1982wi} and ekpyrosis \cite{Khoury:2001wf}. Inflation is a period of accelerated expansion following the big bang, characterized by a large Hubble parameter $H$ and an equation of state $w \approx -1$. Ekpyrosis is a period of ultra-slow contraction preceding the big bang, characterized by a small $H$ and $w > 1$.

A distinctive feature of all currently known ekpyrotic models is that, during the ekpyrotic contraction phase, gravitational waves are not amplified. More specifically, the gravitational waves have a blue spectrum, but their quantum state does not become squeezed; hence, they cannot be given a classical interpretation \cite{Tseng:2012qd,Battarra:2013cha}. During the expanding phase, the (classical) scalar curvature perturbations provide a source for gravitational waves at second order in perturbation theory, leading to a small-amplitude gravitational wave background \cite{Baumann:2007zm}. This small-amplitude background is not compatible with a tensor-to-scalar ratio $r \sim 0.2$, as reported by the BICEP2 collaboration \cite{Ade:2014xna}. However, questions have been raised about the BICEP2 claim \cite{Flauger:2014qra} and it will take other ongoing experiments to determine if there really exist any  detectable tensor B-modes.
Hence, in the meantime, it is reasonable to assume the Planck2013 bound on $r$ and continue studying cyclic/ekpyrotic scenarios, given their conceptual advantages regarding a number of important open issues in early universe cosmology (such as the initial conditions and measure problems). Furthermore, it is conceivable that a detectable gravitational wave spectrum can be created in the context of cyclic/ekpyrotic scenarios, {\it e.g.} by phase transitions, topological defects or during the bounce -- these avenues remain to be explored. 

Using the field picture, the ekpyrotic phase can be described by a scalar field, $\phi$, rolling down a steep negative potential
\begin{equation}
V = - V_0 e^{-\sqrt{2\epsilon}\phi} , 
\end{equation}
where $V_0$ is a constant and $\epsilon$ denotes the equation-of-state parameter 
\begin{equation}\label{es}
\epsilon \equiv \frac{3}{2}\left(1 + w  \right)\quad \text{with}\quad
w \equiv \frac{\rho_S}{p_S}\,,
\end{equation} 
where $w$ is the equation of state, $\rho_S$ the energy density, and 
$p_S$ the pressure of the smoothing background component. 

It has been shown \cite{Tseng:2012qd,Battarra:2013cha} that, if there is only a single field in the contracting phase, the (adiabatic) perturbations are not amplified and cannot be the seed of structure in the post-bang universe.  The currently best-understood way around this problem is the entropic mechanism, where pre-bang isocurvature fluctuations are generated by adding a second ekpyrotic field, $\phi_2$ \cite{Notari:2002yc,Finelli:2002we,Lehners:2007ac,Buchbinder:2007ad}. These isocurvature modes are then converted into density perturbations which source structure in the post-bang universe.

A simple example of an action describing the standard ekpyrotic mechanism is
\begin{eqnarray}\label{old_action}
S &=& \int d^4 x \sqrt{-g}\frac{R}{2}\nonumber\\
&-& \int d^4 x \sqrt{-g}\bigg(\frac{1}{2}\partial_{\mu}\phi_1\partial^{\mu}\phi_1+V_1 e^{-c_1\phi_1}\bigg) 
\nonumber\\
&-&\int d^4 x \sqrt{-g}\bigg(\frac{1}{2}\partial_{\mu}\phi_2\partial^{\mu}\phi_2 + V_2 e^{-c_2\phi_2}\bigg),
\end{eqnarray}
where $V_1, V_2, c_1, c_2$ are constants and the two fields have separate ekpyrotic potentials. (Here and throughout this paper we choose units such that $M_{\text{Pl}}^2 \equiv 1$, where $M_{\text{Pl}}^2 = (8\pi \mathrm{G})^{-1}$ is the reduced Planck mass and $\mathrm{G}$ is Newton's constant.) The background evolution is determined by the linear combination of these potentials, or equivalently, after performing a rotation in field space,  by the adiabatic field, $\sigma$, (defined to point tangentially along the background trajectory, with $\dot{\sigma} = (\dot{\phi_1}^2+\dot{\phi_2}^2)^{1/2}$) while the evolution of perturbations is governed by the entropy field, $s$ (which is, by definition, perpendicular to the $\sigma$-field). At the end of the ekpyrotic phase and before the bounce, the background trajectory bends and  the isocurvature perturbations are converted into adiabatic ones.

However, it is well-known that these ekyprotic solutions for $\phi_1$ and $\phi_2$ are {\it unstable}, in that the $\sigma$ direction runs along a ridge in the potential that is unstable to variations in the $s$ direction (possible consequences in a cyclic context were discussed in \cite{Lehners:2009eg,Lehners:2011ig}).
Also, to obtain nearly scale-invariant spectra requires a steep negative potential which results in the generation of non-negligible non-Gaussianity {\it during the ekpyrotic phase} that dominates the non-Gaussianity generated during the conversion of entropic fluctuations to curvature fluctuations after the ekpyrotic phase \cite{Koyama:2007if,Buchbinder:2007at,Lehners:2007wc,Lehners:2008my}.
Furthermore, the steepness of the potential and the instability involve additional tuning of parameters and initial conditions such that, from a theoretical point of view, it would be desirable to find an alternative approach that avoids them.

In this paper, we explore a new type of entropic mechanism in which there are two scalar fields, as before, but only one has a steep negative potential, $V(\phi)$.  This first field, $\phi$, dominates the energy density and is the source of the ekpyrotic equation of state.  The second field, $\chi$, has a negligible potential, perhaps precisely zero potential, but its kinetic energy density is multiplied by a function of the first field, $\Omega^2(\phi)$, with a non-linear sigma-model type interaction.  This model shows certain similarities with conformal cosmology \cite{Rubakov:2009np} and pseudo-conformal cosmology \cite{Hinterbichler:2011qk}. 

A specific example of our model was introduced in \cite{Li:2013hga} and \cite{Qiu:2013eoa} where both the potential and the non-trivial kinetic coupling are proportional to $e^{-\lambda\phi}$, where $\lambda$ is a positive constant. This model, which is characterized by a constant equation of state $\epsilon$, admits stable scaling solutions that generate (nearly) scale-invariant spectra and, as shown by
\cite{Fertig:2013kwa}, the bispectrum of this model vanishes such that no non-Gaussianity is produced during the ekpyrotic phase.  
As such, these models fit well within the Planck2013 bounds on non-Gaussianity; hence it is worthwhile studying how general these results are.

Here, we show that these results can be extended to an entire class of ekpyrotic models: we show that scale-invariant entropic perturbations can be produced continuously as modes leave the horizon for {\it any time-dependent} ekpyrotic background equation of state. This has the additional advantage of reducing fine-tuning constraints. The corresponding background solutions are stable and the bispectrum of these perturbations vanishes, such that no non-Gaussianity is produced during the ekpyrotic phase. Hence, the only contribution to non-Gaussianity comes from the non-linearity of the conversion process during which entropic perturbations are turned into adiabatic ones. 

The paper is organized as follows. In Sec.\,2 we introduce a generic action involving two fields, derive the background equations of motions and briefly discuss their properties.
In Sec.\,3 we derive the equations of motion at first order in perturbation theory and show that for each background potential, $V(\phi)$, we can define a non-trivial field-space metric such that the spectrum of entropy perturbations, produced by the $\chi$-field, is scale-invariant. We illustrate our finding on a simple class of ekpyrotic models with equation-of-state parameter $\epsilon = \bar{\epsilon}(-\tau)^p$, where $p>0$. 
In Sec.\,4 we compute the bispectrum of the perturbations and we show that, for models with constant spectral tilt, no non-Gaussianity is generated during the ekpyrotic phase. We conclude in Sec.\,5 by summarizing our results and discussing directions for future research.

\section{Setup}

We shall consider the following action involving two scalar fields and a non-trivial field-space metric,
\begin{eqnarray}\label{action}
S &=& \int d^4 x \sqrt{-g}\frac{R}{2} \\
&-& \int d^4 x \sqrt{-g}\bigg(\frac{1}{2}\partial_{\mu}\phi\partial^{\mu}\phi - V(\phi) +  \frac{1}{2}\Omega^2(\phi)\partial_{\mu}\chi\partial^{\mu}\chi\bigg).\nonumber
\end{eqnarray}
Here we work in units where the reduced Planck mass $M_{\textsc{Pl}}^{-2} = 8\pi G = 1$.
With a steep negative potential $V(\phi)$, the first field, $\phi$, dominates the energy density and is the source of the ekpyrotic equation of state. We are assuming that the potential does not depend on the second field, $\chi$; however, the kinetic term of $\chi$ is multiplied by a function of the first field, $\Omega^2(\phi)$, with a non-linear sigma-model type interaction. 
Varying the action with respect to the metric and the fields leads to the equations of motion
\begin{eqnarray}\label{fieldeq1}
H^2 = \frac{1}{3}\Bigg( \frac{1}{2}\dot{\phi}^2 + \frac{1}{2}\Omega^2(\phi)\dot{\chi}^2 + V(\phi) \Bigg)&,&\\\label{fieldeq2}
\ddot{\phi} + 3H \dot{\phi}  - \Omega\,\Omega,_{\phi}\dot{\chi}^2 + V,_{\phi} = 0&,&\\\label{fieldeq3}
\ddot{\chi} +  \left(3H + 2\frac{\dot{\Omega}}{\Omega}  \right)  \dot{\chi} = 0 \label{chieq}&,&
\end{eqnarray}
where we have assumed a flat Friedmann-Lemaitre-Robertson-Walker universe, with $H = \dot{a}/a$ being the Hubble parameter, $a$ the scale factor, and a dot denoting differentiation with respect to physical time $t$.

The crucial ingredient of our model is the non-trivial field-space metric combined with negligible mass of the $\chi$-field: it is immediately apparent that $\dot{\chi} = 0$ is a solution of Eq.~(\ref{chieq}) --  the non-canonical kinetic coupling acts as an additional
friction term,  ``freezing'' $\chi$ as long as $3H + 2\dot{\Omega}/\Omega >0$. 
Having no or negligible potential, the $\chi$ direction is automatically perpendicular to the $\phi$ direction in scalar field space. Hence, the $\dot{\chi} = 0$ solution naturally defines $\chi$ as the entropy field generating first-order entropy/isocurvature fluctuations while $\phi$ remains the adiabatic field controlling the background evolution. By a standard stability analysis, it can easily be shown that the scale-invariant $(\Omega^2, V)$ solutions for $\phi$ and $H$ that we shall discuss below are stable -- a more detailed stability analysis is provided in the Appendix.

\section{Scale-invariance}

Next, we shall show that for an arbitrary ekpyrotic potential $V(\phi)$ there is a non-canonical kinetic coupling $\Omega^2(\phi)$ such that the corresponding spectrum of  entropy perturbations is scale-invariant.

\subsection{The general case}

In order to derive the equations of motion at first order in perturbation theory, we vary the second-order action
\begin{equation} \label{actiondeltachi}
S = \int d^4 \sqrt{-g}\, \Omega^2(\phi) \partial^{\mu}\delta \chi\partial_{\mu}\delta\chi
\end{equation}
with respect to the entropy field perturbation $\delta\chi$.
With the canonically normalized variable $v_s \equiv a\delta s$, where $\delta s \equiv \Omega \delta\chi $ is the gauge invariant entropy perturbation, the linearized equation of motion reads (in Fourier-space and using conformal time $\tau$) \cite{DiMarco:2002eb}
\begin{equation}\label{sv}
v_s'' + \left( k^2 - \frac{\Omega''}{\Omega} - 2\frac{a'}{a}\frac{\Omega'}{\Omega} - \frac{a''}{a}\right)v_s = 0.
\end{equation}
Here, $k$ denotes the wavenumber of the fluctuation mode; and $\tau$ runs from large negative to small negative values during contraction with $\tau_{\textnormal{end}}$ marking the end of ekpyrosis; and a prime denotes a derivative with respect to conformal time.

Assuming standard Bunch-Davies initial conditions, {\it i.e.}, $v_s \rightarrow e^{-ik\tau}/\sqrt{2k}$ for $k\tau \rightarrow -\infty$, the solution of Eq.~(\ref{sv}) is
\begin{equation}
v_s = \sqrt{\frac{\pi}{4}(-\tau)}H_{\nu}^{(1)}(-k\tau),
\end{equation}
where $H_{\nu}^{(1)}$ is a Hankel function of the first kind and $\nu$ is given by
 \begin{equation}\label{nu}
\nu^2 = \frac{1}{4} + \tau^2 \left(\frac{\Omega''}{\Omega} + 2\frac{a'}{a}\frac{\Omega'}{\Omega} + \frac{a''}{a} \right).
\end{equation}
In the late-time/large-scale approximation $v_s$ reduces to  
\begin{equation}\label{v_approx}
v_s \propto k^{-\nu}(-\tau)^{1/2 - \nu}.
\end{equation}
Thus, the spectral index is given by
\begin{equation}
n_{S} - 1 = 3 - 2\nu \, .
\end{equation}
If we now use Eq.~(\ref{nu}), we obtain the following condition for scale-invariance ($\nu=3/2$),
\begin{equation}\label{omegaeq}
\Omega''(\tau) + 2\,\mathcal{H}\,\Omega'(\tau) + \left( \mathcal{H}^2 + \mathcal{H}' - \frac{2}{\tau^2}  \right)\Omega(\tau) = 0,
\end{equation}
where we introduced the conformal Hubble parameter
\begin{equation}
\mathcal{H} = a'(\tau)/a(\tau).
\end{equation}
Eq.~(\ref{omegaeq}) is a homogeneous second-order linear differential equation. Hence, for all continuous $\mathcal{H}$ and all $\tau < \tau_{\textnormal{end}}$ there exists (at least locally) a function $\Omega(\tau)$ such that the resulting spectrum of entropy perturbations is scale-invariant. A global solution exists if the solution $\phi(\tau)$ is a $\mathcal{C}^1$-diffeomorphism, {\it i.e.}, continuously differentiable and invertible. Note that one can repeat straightforwardly this calculation for different values of the spectral index, by choosing an appropriate value for $\nu$ in Eq. (\ref{nu}).

\subsection{An example}

\begin{figure*}%[h]
\begin{center}
\includegraphics[scale=0.475]{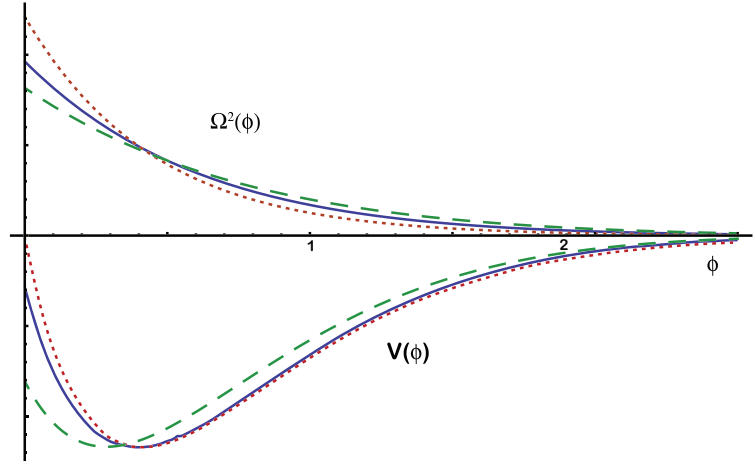}
\caption{A plot of the ekpyrotic potential $V(\phi)$ as in Eq.~(\ref{v_phi}) corresponding to the equation-of-state parameter $\epsilon \equiv \bar{\epsilon}\tau^p$ and the non-canonical kinetic coupling $\Omega(\phi)$ as in Eq.~(\ref{omega_phi}) that together with $V(\phi)$ yields a scale-invariant spectrum, as a function of the ekpyrotic field $\phi$ and for different values of $p$; $p = 0.1$ (dotted red curve), $p=0.5$ (solid blue curve), and $p=0.9$ (dashed green curve). The horizontal axis is in Planck units and the vertical axis uses arbitrary units.} 
\end{center}
\end{figure*}

To illustrate the above analysis we consider ekpyrotic models with a power-law equation-of-state parameter
\begin{equation}\label{eqofstate}
\epsilon \equiv \bar{\epsilon}(-\tau)^p ,\quad 0<p<1,
\end{equation}
where $\epsilon>3, \bar{\epsilon}=$ constant. 
In \cite{Fertig:2013kwa}, the $p=0$ case was considered in which $\epsilon = \bar{\epsilon}$ is constant and where it was assumed that the potentials have some bend or cut-off at $\tau_{\text{end}}$ to reduce $\epsilon$ below 3 and thus end the ekpyrotic phase. Here, for ease of comparison with the constant $\epsilon$ case, we will do the same, taking $\tau_{\text{end}} = - 1$ so that $\epsilon \rightarrow \bar{\epsilon} =$ constant at the end of the ekpyrotic phase (and the potential has a bend or cut-off, as before).

From the second Friedmann equation, $\epsilon = 1 - \mathcal{H}'/\mathcal{H}^2$, we first get  
\begin{equation}\label{scriptH}
\mathcal{H}^{-1} = - \int^{\tau_{\text{end}}}_{\tau}(\epsilon - 1)\, \text{d}\tau ={\tau \left( \frac{\bar{\epsilon}}{p+1}(-\tau)^p - 1 \right)},
\end{equation}
$|\mathcal{H}^{-1}(\tau_{\text{end}})| \ll |\mathcal{H}(\tau)^{-1}|$.
Substituting the expression for $\mathcal{H}$ into Eq.~(\ref{omegaeq}) yields
\begin{eqnarray}\label{omegatau}
\Omega(\tau) &=& \left( \bar{\epsilon} -1 \right)^{1/p} \left(  \frac{\bar{\epsilon}}{p+1}(-\tau)^p - 1 \right)^{-1/p}\nonumber\\ 
&\times&\exp\left( -\frac{\bar{\epsilon}}{\bar{\epsilon} -1} \right),
\end{eqnarray}
where we defined the constants of integration such that $\Omega(\tau)$ corresponds to the constant $\epsilon$ solution for $p \rightarrow 0$.

The expression for the potential is given by the first Friedmann equation, 
%V(tau)
\begin{eqnarray}\label{vtau}
V(\tau) & = & - \frac{\mathcal{H}^2}{a^2}\left(\epsilon - 3 \right)\nonumber\\
&=&  - (p+1)^2 \frac{(\bar{\epsilon} -p- 1)^{2/p} \left(\bar{\epsilon}\,(-\tau)^p - 3\right)}{\left(  \bar{\epsilon}(-\tau)^p - p- 1 \right)^{2+2/p}},
\end{eqnarray}
with
\begin{eqnarray}
a(\tau) &=& a(\tau_{\text{end}}) \exp \left( \int_{\tau}^{\tau_{\text{end}}}\mathcal{H}\, \text{d}\tau \right)\nonumber\\ &=&
\frac{1}{(-\tau)} \left( \frac{\bar{\epsilon} (-\tau)^p -p-1}{\bar{\epsilon} -p-1} \right)^{1/p} ,
\end{eqnarray}
from Eq.~(\ref{scriptH}), and $a(\tau_{\text{end}})$ is an arbitrary constant which we set to unity here.

Next, we want to find an expression for $V$ and $\Omega$ as a function of $\phi$.
Again, we use the second Friedmann equation and find
\begin{eqnarray}\label{phitau}
\phi(\tau) & = & \int_{\tau}^{\tau_\text{end}} \text{d}\tau \sqrt{2\epsilon}\,{\mathcal{H}} \\
&=& \sqrt{2}(p+1) \int_{\tau}^{\tau_\text{end}} \text{d}\tau \frac{\sqrt{\bar{\epsilon}(-\tau)^p}}{\tau \left(\bar{\epsilon}(-\tau)^p - p - 1 \right)} \nonumber\\
&=&  \frac{\sqrt{2}\sqrt{p+1}}{p} \nonumber\\ &\times&
\ln\left( \frac{\sqrt{\bar{\epsilon}}-\sqrt{p+1}}{\sqrt{\bar{\epsilon}} + \sqrt{p+1}}\cdot \frac{\sqrt{\bar{\epsilon}(-\tau)^p} + \sqrt{p+1}}{\sqrt{\bar{\epsilon}(-\tau)^p} - \sqrt{p+1}} \right).\nonumber
\end{eqnarray}
Note that $\phi(\tau) \rightarrow \sqrt{2/\bar{\epsilon}}\ln (-\tau)$ for $p \rightarrow 0$, in agreement with the $\epsilon \equiv \bar{\epsilon}$ solution.
\\
Inverting Eq.~(\ref{phitau}),
\begin{equation}\label{tauphi}
\tau(\phi) = \left( \frac{p+1}{\bar{\epsilon}} \right)^{1/p} \left( \frac{ \frac{\sqrt{\bar{\epsilon}}+\sqrt{p+1}}{\sqrt{\bar{\epsilon}} - \sqrt{p+1}}  \exp \left( \frac{p\phi}{\sqrt{2(p+1)}} \right) +1 }{\frac{\sqrt{\bar{\epsilon}}+\sqrt{p+1}}{\sqrt{\bar{\epsilon}} -\sqrt{p+1}}  \exp \left( \frac{p\phi}{\sqrt{2(p+1)}} \right) - 1} \right)^{2/p},
\end{equation}
and substituting into Eq.~(\ref{omegatau}) allows us to express the kinetic coupling function in terms of the scalar field
%Omega(phi)
\begin{eqnarray}\label{omega_phi}
\Omega(\phi) & =& \exp \left(\frac{-\bar{\epsilon}}{\bar{\epsilon} - 1}  \right) (\bar{\epsilon} - 1)^{1/p}  \left(\frac{\epsilon(\phi)}{p+1} - 1 \right)^{1/p},
\end{eqnarray}
with $\epsilon(\phi)/ (p+1)$ defined as
\begin{eqnarray}
\left(\frac{ (\sqrt{\bar{\epsilon}}+\sqrt{p+1})  \exp \left( \frac{p\phi}{\sqrt{2(p+1)}} \right)+ \sqrt{\bar{\epsilon}} - \sqrt{p+1}
}{(\sqrt{\bar{\epsilon}}+\sqrt{p+1})  \exp \left( \frac{p\phi}{\sqrt{2(p+1)}} \right) - \sqrt{\bar{\epsilon}} + \sqrt{p+1} } \right)^2.
\end{eqnarray}
Note that in the small $p$ limit we recover a simple exponential 
\begin{equation}
\Omega(\phi) \rightarrow   \exp(-\sqrt{\bar{\epsilon}/2}\,\phi) \quad \text{for} \quad p \rightarrow 0. 
\end{equation}
Finally, we can express the potential $V$ as a function of $\phi$. Eq.~(\ref{vtau}) and (\ref{tauphi}) yield
\begin{equation}\label{v_phi}
V(\phi) = - (p+1)^2 (\bar{\epsilon} -p- 1)^{2/p} \frac{\epsilon(\phi) - 3}{\left(  \epsilon(\phi) - p- 1 \right)^{2+2/p}},
\end{equation}
with the small $p$ limit
\begin{equation}
V(\phi) \rightarrow - \frac{\bar{\epsilon}-3}{(\bar{\epsilon}-1)^2} \exp(-\sqrt{2\bar{\epsilon}}\,\phi) \quad \text{for} \quad p \rightarrow 0. 
\end{equation} In particular, we see that for constant equation-of-state, $\Omega^2$ and $V$ need to be proportional to each other in order to yield a scale-invariant spectrum. For examples of scale-invariant $(\Omega^2,V)$ pairs with different values of $p$ see Figure~1. The graph in Fig.~1 shows that both $\Omega(\phi)$ and $V(\phi)$ are simple monotonic functions of $\phi$ and, hence, require no additional tuning whatsoever. This might be surprising since, in general, $\Omega(\phi)$ and $V(\phi)$ do not have simple expressions as a function of $\phi$.  
However, in using a hydrodynamic approach, we start with the assumption that the equation of state takes a simple functional form which, in turn, leads to a simple dynamics. This simplicity is not generally reflected in the field picture in which the equation of state is assumed to derive from a scalar field with canonical kinetic energy.  Conversely, models that may be simple in the field picture may have complicated, time-varying equations of state.  In the analysis here, we are throughout using the hydrodynamic picture and aiming for simplicity within this purely hydrodynamic prescription.
 
As should be clear from the above discussion, this construction works equally well for the case where the spectral index is constant yet different from exact scale-invariance. All one needs to do is choose a different value for the index $\nu$ in Eq. (\ref{nu}), which leads to a slight modification of Eq. (\ref{omegaeq}). For example, repeating the analysis of the present section for the case where we have a deviation from scale-invariance $n_S-1 \equiv -\delta$, leads to a kinetic coupling
\begin{equation}
\Omega \propto \tau^{-\delta/2} \left(  \frac{\bar{\epsilon}(-\tau)^p}{p+1} - 1 \right)^{-1/p}.
\end{equation}

\section{Non-Gaussianity from the ekpyrotic phase}

In the following we show that with scale-invariant $(\Omega^2, V)$ pairs, as introduced in the previous section, no non-Gaussianity is produced during the ekpyrotic phase in the sense that the bispectrum of the perturbations vanishes exactly. Hence, the only contribution to non-Gaussianity comes from the conversion process which is the subdominant contribution in standard ekpyrotic/cyclic theory \cite{Lehners:2007wc,Lehners:2013cka}. We will also extend this result to   $(\Omega^2, V)$ pairs with constant spectral tilt different from $1$.

\subsection{Non-Gaussianity from the ekpyrotic phase}
 
The standard (phenomenological) parameterization of non-Gaussianities is by way of introducing a non- linear correction to a Gaussian perturbation, $\zeta_G$,
\begin{equation}
\zeta(\mathbf{x}) = \zeta_G(\mathbf{x}) + \frac{3}{5}f_{\textsc{nl}}^{\text{loc.}}\left[ \zeta_G^2(\mathbf{x}) - \langle\zeta_G^2(\mathbf{x})\rangle \right].
\end{equation}
This definition is local in real space and thus $f_{\textsc{nl}}^{\text{loc.}}$ is called non-Gaussianity of the {\it local} type. 
 
More generally, the leading non-Gaussian correction is given by the 3-point correlation function, or its Fourier-equivalent, the bispectrum
 \begin{equation}
\langle \zeta_{\mathbf{k}_1}\zeta_{\mathbf{k}_2}\zeta_{\mathbf{k}_3} \rangle= B_{\zeta}(\mathbf{k}_1, \mathbf{k}_2, \mathbf{k}_3).
\end{equation}
For perturbations around an FLRW background, the momentum dependence of the bispectrum simplifies considerably. Homogeneity, or translation invariance, means that the bispectrum must be proportional to a delta function of the sum of the momenta, $B_{\zeta}(\mathbf{k}_1, \mathbf{k}_2, \mathbf{k}_3) \propto \delta(\mathbf{k}_1+ \mathbf{k}_2 + \mathbf{k}_3)$, {\it i.e.}, the sum of the momentum 3-vectors must form a closed triangle. Isotropy, or rotational invariance, dictates that the bispectrum only depends on the magnitudes of the momentum vectors, but not on their orientations,
 \begin{equation}
 B_{\zeta}(\mathbf{k}_1, \mathbf{k}_2, \mathbf{k}_3) = (2\pi)^3\delta(\mathbf{k}_1+ \mathbf{k}_2 + \mathbf{k}_3)B_{\zeta}(k_1, k_2, k_3).
\end{equation}

Different types of non-Gaussianities are described by different shapes of the closed triangle formed by their three momenta \cite{Babich:2004gb}. For $f_{\textsc{nl}}^{\text{loc.}}$ the triangle is ``squeezed,'' {\it i.e.}, $k_1 \ll k_2 \sim k_3$. Here we have ordered the momenta such that $k_1 \leq k_2\leq k_3$.
Higher-derivative interactions can lead to large non-Gaussianities. A key feature of such interactions is that they are suppressed when any individual mode is far outside the horizon. Hence, the bispectrum arising from higher-derivative interactions peaks when all three modes have wavelengths equal to the horizon size, {\it i.e.}, the triangle has a shape $k_1 = k_2 = k_3$, generating non-Gaussianity of the {\it equilateral} type, $f_{\textsc{nl}}^{\text{equil.}}$. 
A shape that is orthogonal to both the local and equilateral templates is called non-Gaussianity of the {\it orthogonal} type, $f_{\textsc{nl}}^{\text{ortho}.}$. This shape also arises in the presence of higher-derivative interactions.

In the absence of higher-derivative kinetic terms in the action as in Eq.~(\ref{action}), no non-Gaussianity of the equilateral or orthogonal type is produced, see {\it e.g.} \cite{Langlois:2008qf}. Therefore, we will focus on the 3-point function of local shape.

During the ekpyrotic phase, non-Gaussianities of the local type can be generated in two ways, either by second-order entropy perturbations, $\delta s^{(2)}$, ({\it intrinsic} non-Gaussianity), or by first-order entropy perturbations, $\delta s^{(1)}$, that source second-order curvature perturbations, $\zeta^{(2)}$, \cite{Lehners:2008my}.  Here, we indicate the perturbative order by a superscript.

At second order and in co-moving gauge, using the methods of \cite{RenauxPetel:2008gi}, for the perturbation in the fields $\phi$ and $\chi$ we find 
\begin{eqnarray}
\delta\phi^{(2)} &= & \frac{1}{2}\delta s^{(1)} \left( \frac{\Omega,_{\phi}}{\Omega} \delta s^{(1)} - \frac{\delta s^{(1)}{}'}{\phi'} \right),\label{deltaphi}
\\
\delta\chi^{(2)} &= & \Omega^{-1}\delta s^{(2)}.
\end{eqnarray}
Since the $\chi$-field is massless and ``frozen'' at background level, there is no source term for the second-order entropy perturbation, $\delta s^{(2)}$, and, hence, no {\it intrinsic} non-Gaussianity is generated during the ekpyrotic phase, in analogy with the constant-$\epsilon$ case treated in \cite{Fertig:2013kwa}. 

In order to calculate the non-Gaussianity in the second-order curvature perturbations that are sourced by first-order entropy perturbations, we use the following formula for the evolution of the gauge-invariant curvature perturbation $\zeta,$
\begin{equation}
\dot{\zeta} = \frac{2H\delta V}{\dot{\phi}^2-2\delta V}.
\end{equation}
This formula, previously discussed in \cite{Lyth:2004gb,Buchbinder:2007at,Lehners:2009qu}, was  
shown in \cite{Fertig:2013kwa} to also apply to theories with non-canonical kinetic terms and has the remarkable property that it is valid to all orders in perturbation theory. Expanding to second order, and using Eq.~(\ref{deltaphi}), we obtain
\begin{equation} \label{zeta2prime}
\zeta^{(2)}{}' = \frac{\mathcal{H}a^2V,_{\phi}}{\phi'^2}\delta s^{(1)} \left( \frac{\Omega,_{\phi}}{\Omega} \delta s^{(1)} - \frac{\delta s^{(1)}{}'}{\phi'} \right).
\end{equation}
In the late-time/large-scale approximation as in Eq.~(\ref{v_approx}) the expression for $\zeta^{(2)}{}'$ reduces to 
\begin{equation}\label{zeta}
\zeta^{(2)}{}' = \frac{\mathcal{H}a^2V,_{\phi}}{\phi'^3}\left(\frac{v_s}{a}\right)^2\left( \frac{\Omega,_{\tau}}{\Omega} - \frac{1-2\nu}{2\tau} + \mathcal{H} \right).
\end{equation}
For the example where the ekpyrotic background equation of state is given by $\epsilon = \bar{\epsilon}(-\tau)^p,$ we can substite our expressions for $\mathcal{H}$ from Eq.~(\ref{scriptH}) and $\Omega$ from Eq.~(\ref{omegatau}) to obtain
\begin{equation}
\zeta^{(2)}{}' = 0.
\end{equation}
Thus, for this example, one can easily see that the second order curvature perturbation does not get sourced during the ekpyrotic phase, and no non-Gaussianity arises in this manner.

Furthermore, as already mentioned above, repeating the analysis with the same background equation of state but allowing for deviations from exact scale-invariance,  $n_S-1 \equiv -\delta$, from Eq.~(\ref{nu}) we first get 
\begin{equation}
\Omega \propto \tau^{-\delta/2} \left(  \frac{\bar{\epsilon}(-\tau)^p}{p+1} - 1 \right)^{-1/p}.
\end{equation}
Then, substituting into Eq.~(\ref{zeta}) once again yields $\zeta^{(2)}{}' = 0$.
That means, during the ekpyrotic phase no non-Gaussianity is generated even for this broader class of ekpyrotic models with non-zero tilt, {\it e.g.} tilt in accord with cosmic microwave background measurements.

In fact, there exists a simple (though less direct) argument that extends the above results to the entire class of models that we are considering: from the second-order action for $\delta\chi,$ Eq. (\ref{actiondeltachi}), one can easily derive its equation of motion,
\begin{equation}
\nabla^\mu \left(\Omega^2 \partial_\mu (\delta\chi)\right) = 0,
\end{equation}
where $\nabla^{\mu}$ denotes the covariant derivative and $\partial_{\mu}$ the partial derivative with respect to the coordinate $x^{\mu}$.

It is evident that $\delta\chi=$ constant is the relevant solution in our case of interest, which, keeping in mind the definition $\delta s = \Omega \delta\chi,$ immediately implies $\delta s \propto \Omega$ and thus 
\begin{equation}
\frac{(\delta s)^\prime}{\delta s} = \frac{\Omega^\prime}{\Omega} = \frac{\Omega_{,\phi}}{\Omega}\phi^\prime,
\end{equation}
from which $\zeta^{(2)}{}' = 0$ follows upon inspection of Eq. (\ref{zeta2prime}). Thus, remarkably, none of our ($\Omega^2, V$) pairs generate any non-Gaussianity during the ekpyrotic phase. One can trace this result back to the fact that the scalar potential does not depend on the entropy field $\chi$.

\subsection{Non-Gaussianity from the conversion process}

Cosmic microwave background experiments measure curvature perturbations $\zeta$, {\it i.e.}, local perturbations in the scale factor that are described by the perturbed metric
\begin{equation}
ds^2 = - dt^2 + a^2(t) e^{2\zeta(t,x^i)}dx^idx_i \, .
\end{equation}
Like in previous entropic models with canonical kinetic terms, we are assuming that after the ekpyrotic phase comes to an end the entropy perturbations get converted into curvature perturbations. Various concrete mechanisms via which this can happen are known: for example, the conversion process can occur right after the ekpyrotic phase (as in \cite{Lehners:2006pu}), or during the bounce itself (see \cite{Battefeld:2007st}). As discussed in \cite{Fertig:2013kwa}, the most important factor determining the final amplitude of local non-Gaussianity is the efficiency of the conversion process. If the conversion is efficient, meaning that the curvature perturbations acquire an amplitude similar to that of the entropy perturbations that source them, then the local non-Gaussianity parameter $f_{\textsc{nl}}$ is expected to be of ${\cal O}(1).$ Concrete numerical calculations support this estimate, with typical values of $f_{\textsc{nl}} \sim 5$ being found \cite{Lehners:2009ja}. It is interesting that such values are in agreement with current measurements by the Planck satellite, yet are in a range that makes them detectable in the future. We note that in contrast to ekpyrotic models with canonical kinetic terms where the contribution to non-Gaussianity from the conversion process is subdominant, in the theory presented here the sole contribution comes from the conversion.

\section{Discussion}

In this paper, we explored a new class of two-field ekpyrotic models with a massive ekpyrotic field governing the background evolution and a second field with no or negligible mass and non-canonical kinetic term. The crucial ingredient of our model is the non-trivial coupling of the background field to the kinetic term of the second, massless field, which plays the role of the entropy field governing the perturbations. Remarkably, we have found that for each background equation of state there exists a non-trivial kinetic coupling such that our model admits scale-invariant solutions (or, more generally, constant spectral index solutions) at first order in perturbation theory.

At second order, we have found that the bispectrum of these perturbations vanishes, such that no non-Gaussianity is produced during the ekpyrotic phase. Hence, the only contribution to non-Gaussianity comes from the non-linearity of the conversion process during which entropic perturbations are turned into adiabatic ones.  This process is model-dependent, but for an efficient conversion mechanism the final bispectrum remains small, with $f_{\textsc{nl}}^{\text{local}} \sim 5$, which is in accord with current cosmic microwave background measurements \cite{Ade:2013ydc}.

This analysis leaves many avenues for future work. A natural extension of our analysis is the calculation of the 4-point function and predictions for the trispectrum (thus extending the analysis of \cite{Lehners:2009ja} to non-trivial field space metrics), in particular since forthcoming data releases from the Planck satellite and large-scale structure experiments will be able to constrain the primordial trispectrum increasingly tightly. Throughout our analysis, we worked with a minimal extension of the standard ekpyrotic theory, studying a two-field Lagrangian. It might be worthwhile to see if a multi-field generalization adds to our model in improving cyclic theories. Similarly, it would be interesting to explore the implications of including a non-negligible mass for the entropy field.

\begin{acknowledgments}
This research was partially supported by the U.S. Department of Energy 
under grant number DE-FG02- 91ER40671 (PJS). 
AI and JLL gratefully acknowledge the support of the European Research Council in the form of the Starting Grant No. 256994 ``StringCosmOS."
The work of AI is supported in part by a grant from the John Templeton Foundation. The opinions expressed in this publication are those of the authors and do not necessarily reflect the views of the John Templeton Foundation.  AI thanks the Physics Department of Princeton University and JLL thanks the Princeton Center for Theoretical Science for hospitality while this research was completed.

\end{acknowledgments}

\appendix
\section{Stable ekpyrotic solutions}

We are interested in the stability of solutions ($\phi,\chi$) with $\dot{\chi} \equiv 0$ since these solutions automatically define the field $\chi$ as the entropy field that generates the isocurvature perturbations and $\phi$ as the ekpyrotic field that governs the background evolution.

For the stability analysis, it is useful to rewrite the field equations Eq.~(\ref{fieldeq1}--\ref{fieldeq3}) in terms of the new variables
\begin{equation}
x= \frac{\dot{\phi}}{\sqrt{6} H},\quad y= \frac{\dot{\chi}}{\sqrt{6}H}, \quad z= \frac{\sqrt{V}}{\sqrt{3} H},\quad \text{and} \quad N = \ln a.%d\tau = a^{-1}d t 
\end{equation}
With these variables we have the autonomous system
\begin{eqnarray}\label{au1}
x,_N &=& 3 (x^2 + y^2 - 1)\left( x + \frac{V,_{\phi}}{\sqrt{6}V}\right) + \frac{\Omega,_{\phi}}{\Omega} y^2  ,\\ \label{au2}
y,_N &=& \left( 3 (x^2 + y^2 - 1)  - \sqrt{6}\frac{\Omega,_{\phi}}{\Omega}x \right)y.
\end{eqnarray}

Next we shall study the behavior of small perturbations around solutions $(x_0, y_0\equiv 0)$. At linear order, the perturbations $\delta x = x - x_0$ and $\delta y = y - y_0$ satisfy the equations 
\begin{eqnarray}\label{pert1}
\delta x,_N & \simeq & \left(9x_0^2 -3 + \sqrt{6}x_0 \left.\frac{V,_{\phi}}{V}\right\vert_{x_0} \right)\delta x,\\ \label{pert2}
\delta y,_N & \simeq & \left( 3x_0^2 - 3  - \sqrt{6} x_0 \left.\frac{\Omega,_{\phi}}{\Omega}\right\vert_{x_0}\right)\delta y,
\end{eqnarray}
where we assumed that $\left.\frac{V,_{\phi}}{V}\right\vert_{x_0}\gg \delta\left(\frac{V,_{\phi}}{V}\right)$.

Since $d N$ is negative during the ekpyrotic phase, a solution $(x_0, 0)$ is stable iff
\begin{eqnarray}\label{sol1}
9x_0^2 -3 &>& - \sqrt{6}x_0 \left.\frac{V,_{\phi}}{V}\right\vert_{x_0},\\ \label{sol2}
 3 (x_0^2 - 1)  &>& \sqrt{6}x_0\left.\frac{\Omega,_{\phi}}{\Omega}\right\vert_{x_0}.
 \end{eqnarray}
If $y_0 = 0$, $x_0 = \sqrt{2\epsilon_0/6}$, where $\epsilon_0$ is the equation-of-state parameter corresponding to $(x_0, 0)$. Using this relation, the stability criteria Eq.~(\ref{sol1}--\ref{sol2}) can be rewritten in terms of the equation-of-state parameter $\epsilon_0$,
\begin{eqnarray}
\frac{\epsilon_0 - 3}{\sqrt{2\epsilon_0}} + \left.\frac{V,_{\phi}}{V}\right\vert_{x_0}&>&0,\\
\frac{3\epsilon_0 - 3}{\sqrt{2\epsilon_0}} - \left.\frac{\Omega,_{\phi}}{\Omega}\right\vert_{x_0}&>&0.
\end{eqnarray}
In particular, $(x_0 = \sqrt{\bar{\epsilon}(-\tau)^p/3}, y_0=0)$, as defined in Eq.~(\ref{eqofstate}), is a stable solution.

\bibliographystyle{apsrev4-1}

\bibliography{ekpyrotic_ng}

\begin{thebibliography}{38}%
\makeatletter
\providecommand \@ifxundefined [1]{%
 \@ifx{#1\undefined}
}%
\providecommand \@ifnum [1]{%
 \ifnum #1\expandafter \@firstoftwo
 \else \expandafter \@secondoftwo
 \fi
}%
\providecommand \@ifx [1]{%
 \ifx #1\expandafter \@firstoftwo
 \else \expandafter \@secondoftwo
 \fi
}%
\providecommand \natexlab [1]{#1}%
\providecommand \enquote  [1]{``#1''}%
\providecommand \bibnamefont  [1]{#1}%
\providecommand \bibfnamefont [1]{#1}%
\providecommand \citenamefont [1]{#1}%
\providecommand \href@noop [0]{\@secondoftwo}%
\providecommand \href [0]{\begingroup \@sanitize@url \@href}%
\providecommand \@href[1]{\@@startlink{#1}\@@href}%
\providecommand \@@href[1]{\endgroup#1\@@endlink}%
\providecommand \@sanitize@url [0]{\catcode `\\12\catcode `\$12\catcode
  `\&12\catcode `\#12\catcode `\^12\catcode `\_12\catcode `\%12\relax}%
\providecommand \@@startlink[1]{}%
\providecommand \@@endlink[0]{}%
\providecommand \url  [0]{\begingroup\@sanitize@url \@url }%
\providecommand \@url [1]{\endgroup\@href {#1}{\urlprefix }}%
\providecommand \urlprefix  [0]{URL }%
\providecommand \Eprint [0]{\href }%
\providecommand \doibase [0]{http://dx.doi.org/}%
\providecommand \selectlanguage [0]{\@gobble}%
\providecommand \bibinfo  [0]{\@secondoftwo}%
\providecommand \bibfield  [0]{\@secondoftwo}%
\providecommand \translation [1]{[#1]}%
\providecommand \BibitemOpen [0]{}%
\providecommand \bibitemStop [0]{}%
\providecommand \bibitemNoStop [0]{.\EOS\space}%
\providecommand \EOS [0]{\spacefactor3000\relax}%
\providecommand \BibitemShut  [1]{\csname bibitem#1\endcsname}%
\let\auto@bib@innerbib\@empty
%</preamble>
\bibitem [{\citenamefont {Ade}\ \emph {et~al.}(2013{\natexlab{a}})\citenamefont
  {Ade} \emph {et~al.}}]{Ade:2013lta}%
  \BibitemOpen
  \bibfield  {author} {\bibinfo {author} {\bibfnamefont {P.}~\bibnamefont
  {Ade}} \emph {et~al.} (\bibinfo {collaboration} {Planck Collaboration}),\
  }\href@noop {} {\  (\bibinfo {year} {2013}{\natexlab{a}})},\ \Eprint
  {http://arxiv.org/abs/1303.5076} {arXiv:1303.5076 [astro-ph.CO]} \BibitemShut
  {NoStop}%
%%CITATION = ARXIV:1303.5076;%%
\bibitem [{\citenamefont {Ade}\ \emph {et~al.}(2013{\natexlab{b}})\citenamefont
  {Ade} \emph {et~al.}}]{Ade:2013rta}%
  \BibitemOpen
  \bibfield  {author} {\bibinfo {author} {\bibfnamefont {P.}~\bibnamefont
  {Ade}} \emph {et~al.} (\bibinfo {collaboration} {Planck Collaboration}),\
  }\href@noop {} {\  (\bibinfo {year} {2013}{\natexlab{b}})},\ \Eprint
  {http://arxiv.org/abs/1303.5082} {arXiv:1303.5082 [astro-ph.CO]} \BibitemShut
  {NoStop}%
%%CITATION = ARXIV:1303.5082;%%
\bibitem [{\citenamefont {Ade}\ \emph {et~al.}(2013{\natexlab{c}})\citenamefont
  {Ade} \emph {et~al.}}]{Ade:2013ydc}%
  \BibitemOpen
  \bibfield  {author} {\bibinfo {author} {\bibfnamefont {P.}~\bibnamefont
  {Ade}} \emph {et~al.} (\bibinfo {collaboration} {Planck Collaboration}),\
  }\href@noop {} {\  (\bibinfo {year} {2013}{\natexlab{c}})},\ \Eprint
  {http://arxiv.org/abs/1303.5084} {arXiv:1303.5084 [astro-ph.CO]} \BibitemShut
  {NoStop}%
%%CITATION = ARXIV:1303.5084;%%
\bibitem [{\citenamefont {Sievers}\ \emph {et~al.}(2013)\citenamefont {Sievers}
  \emph {et~al.}}]{Sievers:2013ica}%
  \BibitemOpen
  \bibfield  {author} {\bibinfo {author} {\bibfnamefont {J.~L.}\ \bibnamefont
  {Sievers}} \emph {et~al.} (\bibinfo {collaboration} {Atacama Cosmology
  Telescope}),\ }\href {\doibase 10.1088/1475-7516/2013/10/060} {\bibfield
  {journal} {\bibinfo  {journal} {JCAP}\ }\textbf {\bibinfo {volume} {1310}},\
  \bibinfo {pages} {060} (\bibinfo {year} {2013})},\ \Eprint
  {http://arxiv.org/abs/1301.0824} {arXiv:1301.0824 [astro-ph.CO]} \BibitemShut
  {NoStop}%
%%CITATION = ARXIV:1301.0824;%%
\bibitem [{\citenamefont {Guth}(1981)}]{Guth:1980zm}%
  \BibitemOpen
  \bibfield  {author} {\bibinfo {author} {\bibfnamefont {A.~H.}\ \bibnamefont
  {Guth}},\ }\href {\doibase 10.1103/PhysRevD.23.347} {\bibfield  {journal}
  {\bibinfo  {journal} {Phys.Rev.}\ }\textbf {\bibinfo {volume} {D23}},\
  \bibinfo {pages} {347} (\bibinfo {year} {1981})}\BibitemShut {NoStop}%
%%CITATION = PHRVA,D23,347;%%
\bibitem [{\citenamefont {Linde}(1982)}]{Linde:1981mu}%
  \BibitemOpen
  \bibfield  {author} {\bibinfo {author} {\bibfnamefont {A.~D.}\ \bibnamefont
  {Linde}},\ }\href {\doibase 10.1016/0370-2693(82)91219-9} {\bibfield
  {journal} {\bibinfo  {journal} {Phys.Lett.}\ }\textbf {\bibinfo {volume}
  {B108}},\ \bibinfo {pages} {389} (\bibinfo {year} {1982})}\BibitemShut
  {NoStop}%
%%CITATION = PHLTA,B108,389;%%
\bibitem [{\citenamefont {Albrecht}\ and\ \citenamefont
  {Steinhardt}(1982)}]{Albrecht:1982wi}%
  \BibitemOpen
  \bibfield  {author} {\bibinfo {author} {\bibfnamefont {A.}~\bibnamefont
  {Albrecht}}\ and\ \bibinfo {author} {\bibfnamefont {P.~J.}\ \bibnamefont
  {Steinhardt}},\ }\href {\doibase 10.1103/PhysRevLett.48.1220} {\bibfield
  {journal} {\bibinfo  {journal} {Phys.Rev.Lett.}\ }\textbf {\bibinfo {volume}
  {48}},\ \bibinfo {pages} {1220} (\bibinfo {year} {1982})}\BibitemShut
  {NoStop}%
%%CITATION = PRLTA,48,1220;%%
\bibitem [{\citenamefont {Khoury}\ \emph {et~al.}(2001)\citenamefont {Khoury},
  \citenamefont {Ovrut}, \citenamefont {Steinhardt},\ and\ \citenamefont
  {Turok}}]{Khoury:2001wf}%
  \BibitemOpen
  \bibfield  {author} {\bibinfo {author} {\bibfnamefont {J.}~\bibnamefont
  {Khoury}}, \bibinfo {author} {\bibfnamefont {B.~A.}\ \bibnamefont {Ovrut}},
  \bibinfo {author} {\bibfnamefont {P.~J.}\ \bibnamefont {Steinhardt}}, \ and\
  \bibinfo {author} {\bibfnamefont {N.}~\bibnamefont {Turok}},\ }\href@noop {}
  {\bibfield  {journal} {\bibinfo  {journal} {Phys. Rev.}\ }\textbf {\bibinfo
  {volume} {D64}},\ \bibinfo {pages} {123522} (\bibinfo {year} {2001})},\
  \Eprint {http://arxiv.org/abs/hep-th/0103239} {hep-th/0103239} \BibitemShut
  {NoStop}%
%%CITATION = HEP-TH/0103239;%%
\bibitem [{\citenamefont {Tseng}(2013)}]{Tseng:2012qd}%
  \BibitemOpen
  \bibfield  {author} {\bibinfo {author} {\bibfnamefont {C.-Y.}\ \bibnamefont
  {Tseng}},\ }\href {\doibase 10.1103/PhysRevD.87.023518} {\bibfield  {journal}
  {\bibinfo  {journal} {Phys.Rev.}\ }\textbf {\bibinfo {volume} {D87}},\
  \bibinfo {pages} {023518} (\bibinfo {year} {2013})},\ \Eprint
  {http://arxiv.org/abs/1210.0581} {arXiv:1210.0581 [hep-th]} \BibitemShut
  {NoStop}%
%%CITATION = ARXIV:1210.0581;%%
\bibitem [{\citenamefont {Battarra}\ and\ \citenamefont
  {Lehners}(2013)}]{Battarra:2013cha}%
  \BibitemOpen
  \bibfield  {author} {\bibinfo {author} {\bibfnamefont {L.}~\bibnamefont
  {Battarra}}\ and\ \bibinfo {author} {\bibfnamefont {J.-L.}\ \bibnamefont
  {Lehners}},\ }\href@noop {} {\  (\bibinfo {year} {2013})},\ \Eprint
  {http://arxiv.org/abs/1309.2281} {arXiv:1309.2281 [hep-th]} \BibitemShut
  {NoStop}%
%%CITATION = ARXIV:1309.2281;%%
\bibitem [{\citenamefont {Baumann}\ \emph {et~al.}(2007)\citenamefont
  {Baumann}, \citenamefont {Steinhardt}, \citenamefont {Takahashi},\ and\
  \citenamefont {Ichiki}}]{Baumann:2007zm}%
  \BibitemOpen
  \bibfield  {author} {\bibinfo {author} {\bibfnamefont {D.}~\bibnamefont
  {Baumann}}, \bibinfo {author} {\bibfnamefont {P.~J.}\ \bibnamefont
  {Steinhardt}}, \bibinfo {author} {\bibfnamefont {K.}~\bibnamefont
  {Takahashi}}, \ and\ \bibinfo {author} {\bibfnamefont {K.}~\bibnamefont
  {Ichiki}},\ }\href {\doibase 10.1103/PhysRevD.76.084019} {\bibfield
  {journal} {\bibinfo  {journal} {Phys. Rev.}\ }\textbf {\bibinfo {volume}
  {D76}},\ \bibinfo {pages} {084019} (\bibinfo {year} {2007})},\ \Eprint
  {http://arxiv.org/abs/hep-th/0703290} {arXiv:hep-th/0703290} \BibitemShut
  {NoStop}%
%%CITATION = HEP-TH/0703290;%%
\bibitem [{\citenamefont {Ade}\ \emph {et~al.}(2014)\citenamefont {Ade} \emph
  {et~al.}}]{Ade:2014xna}%
  \BibitemOpen
  \bibfield  {author} {\bibinfo {author} {\bibfnamefont {P.}~\bibnamefont
  {Ade}} \emph {et~al.} (\bibinfo {collaboration} {BICEP2 Collaboration}),\
  }\href@noop {} {\  (\bibinfo {year} {2014})},\ \Eprint
  {http://arxiv.org/abs/1403.3985} {arXiv:1403.3985 [astro-ph.CO]} \BibitemShut
  {NoStop}%
%%CITATION = ARXIV:1403.3985;%%
\bibitem [{\citenamefont {Flauger}\ \emph {et~al.}(2014)\citenamefont
  {Flauger}, \citenamefont {Hill},\ and\ \citenamefont
  {Spergel}}]{Flauger:2014qra}%
  \BibitemOpen
  \bibfield  {author} {\bibinfo {author} {\bibfnamefont {R.}~\bibnamefont
  {Flauger}}, \bibinfo {author} {\bibfnamefont {J.~C.}\ \bibnamefont {Hill}}, \
  and\ \bibinfo {author} {\bibfnamefont {D.~N.}\ \bibnamefont {Spergel}},\
  }\href@noop {} {\  (\bibinfo {year} {2014})},\ \Eprint
  {http://arxiv.org/abs/1405.7351} {arXiv:1405.7351 [astro-ph.CO]} \BibitemShut
  {NoStop}%
%%CITATION = ARXIV:1405.7351;%%
\bibitem [{\citenamefont {Notari}\ and\ \citenamefont
  {Riotto}(2002)}]{Notari:2002yc}%
  \BibitemOpen
  \bibfield  {author} {\bibinfo {author} {\bibfnamefont {A.}~\bibnamefont
  {Notari}}\ and\ \bibinfo {author} {\bibfnamefont {A.}~\bibnamefont
  {Riotto}},\ }\href {\doibase 10.1016/S0550-3213(02)00765-4} {\bibfield
  {journal} {\bibinfo  {journal} {Nucl.Phys.}\ }\textbf {\bibinfo {volume}
  {B644}},\ \bibinfo {pages} {371} (\bibinfo {year} {2002})},\ \Eprint
  {http://arxiv.org/abs/hep-th/0205019} {arXiv:hep-th/0205019 [hep-th]}
  \BibitemShut {NoStop}%
%%CITATION = HEP-TH/0205019;%%
\bibitem [{\citenamefont {Finelli}(2002)}]{Finelli:2002we}%
  \BibitemOpen
  \bibfield  {author} {\bibinfo {author} {\bibfnamefont {F.}~\bibnamefont
  {Finelli}},\ }\href {\doibase 10.1016/S0370-2693(02)02554-6} {\bibfield
  {journal} {\bibinfo  {journal} {Phys.Lett.}\ }\textbf {\bibinfo {volume}
  {B545}},\ \bibinfo {pages} {1} (\bibinfo {year} {2002})},\ \Eprint
  {http://arxiv.org/abs/hep-th/0206112} {arXiv:hep-th/0206112 [hep-th]}
  \BibitemShut {NoStop}%
%%CITATION = HEP-TH/0206112;%%
\bibitem [{\citenamefont {Lehners}\ \emph
  {et~al.}(2007{\natexlab{a}})\citenamefont {Lehners}, \citenamefont
  {McFadden}, \citenamefont {Turok},\ and\ \citenamefont
  {Steinhardt}}]{Lehners:2007ac}%
  \BibitemOpen
  \bibfield  {author} {\bibinfo {author} {\bibfnamefont {J.-L.}\ \bibnamefont
  {Lehners}}, \bibinfo {author} {\bibfnamefont {P.}~\bibnamefont {McFadden}},
  \bibinfo {author} {\bibfnamefont {N.}~\bibnamefont {Turok}}, \ and\ \bibinfo
  {author} {\bibfnamefont {P.~J.}\ \bibnamefont {Steinhardt}},\ }\href
  {\doibase 10.1103/PhysRevD.76.103501} {\bibfield  {journal} {\bibinfo
  {journal} {Phys.Rev.}\ }\textbf {\bibinfo {volume} {D76}},\ \bibinfo {pages}
  {103501} (\bibinfo {year} {2007}{\natexlab{a}})},\ \Eprint
  {http://arxiv.org/abs/hep-th/0702153} {arXiv:hep-th/0702153 [HEP-TH]}
  \BibitemShut {NoStop}%
%%CITATION = HEP-TH/0702153;%%
\bibitem [{\citenamefont {Buchbinder}\ \emph {et~al.}(2007)\citenamefont
  {Buchbinder}, \citenamefont {Khoury},\ and\ \citenamefont
  {Ovrut}}]{Buchbinder:2007ad}%
  \BibitemOpen
  \bibfield  {author} {\bibinfo {author} {\bibfnamefont {E.~I.}\ \bibnamefont
  {Buchbinder}}, \bibinfo {author} {\bibfnamefont {J.}~\bibnamefont {Khoury}},
  \ and\ \bibinfo {author} {\bibfnamefont {B.~A.}\ \bibnamefont {Ovrut}},\
  }\href {\doibase 10.1103/PhysRevD.76.123503} {\bibfield  {journal} {\bibinfo
  {journal} {Phys.Rev.}\ }\textbf {\bibinfo {volume} {D76}},\ \bibinfo {pages}
  {123503} (\bibinfo {year} {2007})},\ \Eprint
  {http://arxiv.org/abs/hep-th/0702154} {arXiv:hep-th/0702154 [hep-th]}
  \BibitemShut {NoStop}%
%%CITATION = HEP-TH/0702154;%%
\bibitem [{\citenamefont {Lehners}\ \emph {et~al.}(2009)\citenamefont
  {Lehners}, \citenamefont {Steinhardt},\ and\ \citenamefont
  {Turok}}]{Lehners:2009eg}%
  \BibitemOpen
  \bibfield  {author} {\bibinfo {author} {\bibfnamefont {J.-L.}\ \bibnamefont
  {Lehners}}, \bibinfo {author} {\bibfnamefont {P.~J.}\ \bibnamefont
  {Steinhardt}}, \ and\ \bibinfo {author} {\bibfnamefont {N.}~\bibnamefont
  {Turok}},\ }\href {\doibase 10.1142/S0218271809015977} {\bibfield  {journal}
  {\bibinfo  {journal} {Int.J.Mod.Phys.}\ }\textbf {\bibinfo {volume} {D18}},\
  \bibinfo {pages} {2231} (\bibinfo {year} {2009})},\ \Eprint
  {http://arxiv.org/abs/0910.0834} {arXiv:0910.0834 [hep-th]} \BibitemShut
  {NoStop}%
%%CITATION = ARXIV:0910.0834;%%
\bibitem [{\citenamefont {Lehners}(2011)}]{Lehners:2011ig}%
  \BibitemOpen
  \bibfield  {author} {\bibinfo {author} {\bibfnamefont {J.-L.}\ \bibnamefont
  {Lehners}},\ }\href {\doibase 10.1103/PhysRevD.84.103518} {\bibfield
  {journal} {\bibinfo  {journal} {Phys.Rev.}\ }\textbf {\bibinfo {volume}
  {D84}},\ \bibinfo {pages} {103518} (\bibinfo {year} {2011})},\ \Eprint
  {http://arxiv.org/abs/1107.4551} {arXiv:1107.4551 [hep-th]} \BibitemShut
  {NoStop}%
%%CITATION = ARXIV:1107.4551;%%
\bibitem [{\citenamefont {Koyama}\ \emph {et~al.}(2007)\citenamefont {Koyama},
  \citenamefont {Mizuno}, \citenamefont {Vernizzi},\ and\ \citenamefont
  {Wands}}]{Koyama:2007if}%
  \BibitemOpen
  \bibfield  {author} {\bibinfo {author} {\bibfnamefont {K.}~\bibnamefont
  {Koyama}}, \bibinfo {author} {\bibfnamefont {S.}~\bibnamefont {Mizuno}},
  \bibinfo {author} {\bibfnamefont {F.}~\bibnamefont {Vernizzi}}, \ and\
  \bibinfo {author} {\bibfnamefont {D.}~\bibnamefont {Wands}},\ }\href
  {\doibase 10.1088/1475-7516/2007/11/024} {\bibfield  {journal} {\bibinfo
  {journal} {JCAP}\ }\textbf {\bibinfo {volume} {0711}},\ \bibinfo {pages}
  {024} (\bibinfo {year} {2007})},\ \Eprint {http://arxiv.org/abs/0708.4321}
  {arXiv:0708.4321 [hep-th]} \BibitemShut {NoStop}%
%%CITATION = ARXIV:0708.4321;%%
\bibitem [{\citenamefont {Buchbinder}\ \emph {et~al.}(2008)\citenamefont
  {Buchbinder}, \citenamefont {Khoury},\ and\ \citenamefont
  {Ovrut}}]{Buchbinder:2007at}%
  \BibitemOpen
  \bibfield  {author} {\bibinfo {author} {\bibfnamefont {E.~I.}\ \bibnamefont
  {Buchbinder}}, \bibinfo {author} {\bibfnamefont {J.}~\bibnamefont {Khoury}},
  \ and\ \bibinfo {author} {\bibfnamefont {B.~A.}\ \bibnamefont {Ovrut}},\
  }\href {\doibase 10.1103/PhysRevLett.100.171302} {\bibfield  {journal}
  {\bibinfo  {journal} {Phys.Rev.Lett.}\ }\textbf {\bibinfo {volume} {100}},\
  \bibinfo {pages} {171302} (\bibinfo {year} {2008})},\ \Eprint
  {http://arxiv.org/abs/0710.5172} {arXiv:0710.5172 [hep-th]} \BibitemShut
  {NoStop}%
%%CITATION = ARXIV:0710.5172;%%
\bibitem [{\citenamefont {Lehners}\ and\ \citenamefont
  {Steinhardt}(2008{\natexlab{a}})}]{Lehners:2007wc}%
  \BibitemOpen
  \bibfield  {author} {\bibinfo {author} {\bibfnamefont {J.-L.}\ \bibnamefont
  {Lehners}}\ and\ \bibinfo {author} {\bibfnamefont {P.~J.}\ \bibnamefont
  {Steinhardt}},\ }\href {\doibase 10.1103/PhysRevD.79.129903,
  10.1103/PhysRevD.77.063533} {\bibfield  {journal} {\bibinfo  {journal}
  {Phys.Rev.}\ }\textbf {\bibinfo {volume} {D77}},\ \bibinfo {pages} {063533}
  (\bibinfo {year} {2008}{\natexlab{a}})},\ \Eprint
  {http://arxiv.org/abs/0712.3779} {arXiv:0712.3779 [hep-th]} \BibitemShut
  {NoStop}%
%%CITATION = ARXIV:0712.3779;%%
\bibitem [{\citenamefont {Lehners}\ and\ \citenamefont
  {Steinhardt}(2008{\natexlab{b}})}]{Lehners:2008my}%
  \BibitemOpen
  \bibfield  {author} {\bibinfo {author} {\bibfnamefont {J.-L.}\ \bibnamefont
  {Lehners}}\ and\ \bibinfo {author} {\bibfnamefont {P.~J.}\ \bibnamefont
  {Steinhardt}},\ }\href {\doibase 10.1103/PhysRevD.78.023506,
  10.1103/PhysRevD.79.129902} {\bibfield  {journal} {\bibinfo  {journal}
  {Phys.Rev.}\ }\textbf {\bibinfo {volume} {D78}},\ \bibinfo {pages} {023506}
  (\bibinfo {year} {2008}{\natexlab{b}})},\ \Eprint
  {http://arxiv.org/abs/0804.1293} {arXiv:0804.1293 [hep-th]} \BibitemShut
  {NoStop}%
%%CITATION = ARXIV:0804.1293;%%
\bibitem [{\citenamefont {Rubakov}(2009)}]{Rubakov:2009np}%
  \BibitemOpen
  \bibfield  {author} {\bibinfo {author} {\bibfnamefont {V.}~\bibnamefont
  {Rubakov}},\ }\href {\doibase 10.1088/1475-7516/2009/09/030} {\bibfield
  {journal} {\bibinfo  {journal} {JCAP}\ }\textbf {\bibinfo {volume} {0909}},\
  \bibinfo {pages} {030} (\bibinfo {year} {2009})},\ \Eprint
  {http://arxiv.org/abs/0906.3693} {arXiv:0906.3693 [hep-th]} \BibitemShut
  {NoStop}%
%%CITATION = ARXIV:0906.3693;%%
\bibitem [{\citenamefont {Hinterbichler}\ and\ \citenamefont
  {Khoury}(2012)}]{Hinterbichler:2011qk}%
  \BibitemOpen
  \bibfield  {author} {\bibinfo {author} {\bibfnamefont {K.}~\bibnamefont
  {Hinterbichler}}\ and\ \bibinfo {author} {\bibfnamefont {J.}~\bibnamefont
  {Khoury}},\ }\href {\doibase 10.1088/1475-7516/2012/04/023} {\bibfield
  {journal} {\bibinfo  {journal} {JCAP}\ }\textbf {\bibinfo {volume} {1204}},\
  \bibinfo {pages} {023} (\bibinfo {year} {2012})},\ \Eprint
  {http://arxiv.org/abs/1106.1428} {arXiv:1106.1428 [hep-th]} \BibitemShut
  {NoStop}%
%%CITATION = ARXIV:1106.1428;%%
\bibitem [{\citenamefont {Li}(2013)}]{Li:2013hga}%
  \BibitemOpen
  \bibfield  {author} {\bibinfo {author} {\bibfnamefont {M.}~\bibnamefont
  {Li}},\ }\href {\doibase 10.1016/j.physletb.2013.06.035} {\bibfield
  {journal} {\bibinfo  {journal} {Phys.Lett.}\ }\textbf {\bibinfo {volume}
  {B724}},\ \bibinfo {pages} {192} (\bibinfo {year} {2013})},\ \Eprint
  {http://arxiv.org/abs/1306.0191} {arXiv:1306.0191 [hep-th]} \BibitemShut
  {NoStop}%
%%CITATION = ARXIV:1306.0191;%%
\bibitem [{\citenamefont {Qiu}\ \emph {et~al.}(2013)\citenamefont {Qiu},
  \citenamefont {Gao},\ and\ \citenamefont {Saridakis}}]{Qiu:2013eoa}%
  \BibitemOpen
  \bibfield  {author} {\bibinfo {author} {\bibfnamefont {T.}~\bibnamefont
  {Qiu}}, \bibinfo {author} {\bibfnamefont {X.}~\bibnamefont {Gao}}, \ and\
  \bibinfo {author} {\bibfnamefont {E.~N.}\ \bibnamefont {Saridakis}},\ }\href
  {\doibase 10.1103/PhysRevD.88.043525} {\bibfield  {journal} {\bibinfo
  {journal} {Phys.Rev.}\ }\textbf {\bibinfo {volume} {D88}},\ \bibinfo {pages}
  {043525} (\bibinfo {year} {2013})},\ \Eprint {http://arxiv.org/abs/1303.2372}
  {arXiv:1303.2372 [astro-ph.CO]} \BibitemShut {NoStop}%
%%CITATION = ARXIV:1303.2372;%%
\bibitem [{\citenamefont {Fertig}\ \emph {et~al.}(2013)\citenamefont {Fertig},
  \citenamefont {Lehners},\ and\ \citenamefont {Mallwitz}}]{Fertig:2013kwa}%
  \BibitemOpen
  \bibfield  {author} {\bibinfo {author} {\bibfnamefont {A.}~\bibnamefont
  {Fertig}}, \bibinfo {author} {\bibfnamefont {J.-L.}\ \bibnamefont {Lehners}},
  \ and\ \bibinfo {author} {\bibfnamefont {E.}~\bibnamefont {Mallwitz}},\
  }\href@noop {} {\  (\bibinfo {year} {2013})},\ \Eprint
  {http://arxiv.org/abs/1310.8133} {arXiv:1310.8133 [hep-th]} \BibitemShut
  {NoStop}%
%%CITATION = ARXIV:1310.8133;%%
\bibitem [{\citenamefont {Di~Marco}\ \emph {et~al.}(2003)\citenamefont
  {Di~Marco}, \citenamefont {Finelli},\ and\ \citenamefont
  {Brandenberger}}]{DiMarco:2002eb}%
  \BibitemOpen
  \bibfield  {author} {\bibinfo {author} {\bibfnamefont {F.}~\bibnamefont
  {Di~Marco}}, \bibinfo {author} {\bibfnamefont {F.}~\bibnamefont {Finelli}}, \
  and\ \bibinfo {author} {\bibfnamefont {R.}~\bibnamefont {Brandenberger}},\
  }\href {\doibase 10.1103/PhysRevD.67.063512} {\bibfield  {journal} {\bibinfo
  {journal} {Phys.Rev.}\ }\textbf {\bibinfo {volume} {D67}},\ \bibinfo {pages}
  {063512} (\bibinfo {year} {2003})},\ \Eprint
  {http://arxiv.org/abs/astro-ph/0211276} {arXiv:astro-ph/0211276 [astro-ph]}
  \BibitemShut {NoStop}%
%%CITATION = ASTRO-PH/0211276;%%
\bibitem [{\citenamefont {Lehners}\ and\ \citenamefont
  {Steinhardt}(2013)}]{Lehners:2013cka}%
  \BibitemOpen
  \bibfield  {author} {\bibinfo {author} {\bibfnamefont {J.-L.}\ \bibnamefont
  {Lehners}}\ and\ \bibinfo {author} {\bibfnamefont {P.~J.}\ \bibnamefont
  {Steinhardt}},\ }\href {\doibase 10.1103/PhysRevD.87.123533} {\bibfield
  {journal} {\bibinfo  {journal} {Phys.Rev.}\ }\textbf {\bibinfo {volume}
  {D87}},\ \bibinfo {pages} {123533} (\bibinfo {year} {2013})},\ \Eprint
  {http://arxiv.org/abs/1304.3122} {arXiv:1304.3122 [astro-ph.CO]} \BibitemShut
  {NoStop}%
%%CITATION = ARXIV:1304.3122;%%
\bibitem [{\citenamefont {Babich}\ \emph {et~al.}(2004)\citenamefont {Babich},
  \citenamefont {Creminelli},\ and\ \citenamefont
  {Zaldarriaga}}]{Babich:2004gb}%
  \BibitemOpen
  \bibfield  {author} {\bibinfo {author} {\bibfnamefont {D.}~\bibnamefont
  {Babich}}, \bibinfo {author} {\bibfnamefont {P.}~\bibnamefont {Creminelli}},
  \ and\ \bibinfo {author} {\bibfnamefont {M.}~\bibnamefont {Zaldarriaga}},\
  }\href {\doibase 10.1088/1475-7516/2004/08/009} {\bibfield  {journal}
  {\bibinfo  {journal} {JCAP}\ }\textbf {\bibinfo {volume} {0408}},\ \bibinfo
  {pages} {009} (\bibinfo {year} {2004})},\ \Eprint
  {http://arxiv.org/abs/astro-ph/0405356} {arXiv:astro-ph/0405356 [astro-ph]}
  \BibitemShut {NoStop}%
%%CITATION = ASTRO-PH/0405356;%%
\bibitem [{\citenamefont {Langlois}\ \emph {et~al.}(2008)\citenamefont
  {Langlois}, \citenamefont {Renaux-Petel}, \citenamefont {Steer},\ and\
  \citenamefont {Tanaka}}]{Langlois:2008qf}%
  \BibitemOpen
  \bibfield  {author} {\bibinfo {author} {\bibfnamefont {D.}~\bibnamefont
  {Langlois}}, \bibinfo {author} {\bibfnamefont {S.}~\bibnamefont
  {Renaux-Petel}}, \bibinfo {author} {\bibfnamefont {D.~A.}\ \bibnamefont
  {Steer}}, \ and\ \bibinfo {author} {\bibfnamefont {T.}~\bibnamefont
  {Tanaka}},\ }\href {\doibase 10.1103/PhysRevD.78.063523} {\bibfield
  {journal} {\bibinfo  {journal} {Phys.Rev.}\ }\textbf {\bibinfo {volume}
  {D78}},\ \bibinfo {pages} {063523} (\bibinfo {year} {2008})},\ \Eprint
  {http://arxiv.org/abs/0806.0336} {arXiv:0806.0336 [hep-th]} \BibitemShut
  {NoStop}%
%%CITATION = ARXIV:0806.0336;%%
\bibitem [{\citenamefont {Renaux-Petel}\ and\ \citenamefont
  {Tasinato}(2009)}]{RenauxPetel:2008gi}%
  \BibitemOpen
  \bibfield  {author} {\bibinfo {author} {\bibfnamefont {S.}~\bibnamefont
  {Renaux-Petel}}\ and\ \bibinfo {author} {\bibfnamefont {G.}~\bibnamefont
  {Tasinato}},\ }\href {\doibase 10.1088/1475-7516/2009/01/012} {\bibfield
  {journal} {\bibinfo  {journal} {JCAP}\ }\textbf {\bibinfo {volume} {0901}},\
  \bibinfo {pages} {012} (\bibinfo {year} {2009})},\ \Eprint
  {http://arxiv.org/abs/0810.2405} {arXiv:0810.2405 [hep-th]} \BibitemShut
  {NoStop}%
%%CITATION = ARXIV:0810.2405;%%
\bibitem [{\citenamefont {Lyth}\ \emph {et~al.}(2005)\citenamefont {Lyth},
  \citenamefont {Malik},\ and\ \citenamefont {Sasaki}}]{Lyth:2004gb}%
  \BibitemOpen
  \bibfield  {author} {\bibinfo {author} {\bibfnamefont {D.~H.}\ \bibnamefont
  {Lyth}}, \bibinfo {author} {\bibfnamefont {K.~A.}\ \bibnamefont {Malik}}, \
  and\ \bibinfo {author} {\bibfnamefont {M.}~\bibnamefont {Sasaki}},\ }\href
  {\doibase 10.1088/1475-7516/2005/05/004} {\bibfield  {journal} {\bibinfo
  {journal} {JCAP}\ }\textbf {\bibinfo {volume} {0505}},\ \bibinfo {pages}
  {004} (\bibinfo {year} {2005})},\ \Eprint
  {http://arxiv.org/abs/astro-ph/0411220} {arXiv:astro-ph/0411220} \BibitemShut
  {NoStop}%
%%CITATION = ASTRO-PH/0411220;%%
\bibitem [{\citenamefont {Lehners}\ and\ \citenamefont
  {Steinhardt}(2009)}]{Lehners:2009qu}%
  \BibitemOpen
  \bibfield  {author} {\bibinfo {author} {\bibfnamefont {J.-L.}\ \bibnamefont
  {Lehners}}\ and\ \bibinfo {author} {\bibfnamefont {P.~J.}\ \bibnamefont
  {Steinhardt}},\ }\href {\doibase 10.1103/PhysRevD.80.103520} {\bibfield
  {journal} {\bibinfo  {journal} {Phys.Rev.}\ }\textbf {\bibinfo {volume}
  {D80}},\ \bibinfo {pages} {103520} (\bibinfo {year} {2009})},\ \Eprint
  {http://arxiv.org/abs/0909.2558} {arXiv:0909.2558 [hep-th]} \BibitemShut
  {NoStop}%
%%CITATION = ARXIV:0909.2558;%%
\bibitem [{\citenamefont {Lehners}\ \emph
  {et~al.}(2007{\natexlab{b}})\citenamefont {Lehners}, \citenamefont
  {McFadden},\ and\ \citenamefont {Turok}}]{Lehners:2006pu}%
  \BibitemOpen
  \bibfield  {author} {\bibinfo {author} {\bibfnamefont {J.-L.}\ \bibnamefont
  {Lehners}}, \bibinfo {author} {\bibfnamefont {P.}~\bibnamefont {McFadden}}, \
  and\ \bibinfo {author} {\bibfnamefont {N.}~\bibnamefont {Turok}},\ }\href
  {\doibase 10.1103/PhysRevD.75.103510} {\bibfield  {journal} {\bibinfo
  {journal} {Phys.Rev.}\ }\textbf {\bibinfo {volume} {D75}},\ \bibinfo {pages}
  {103510} (\bibinfo {year} {2007}{\natexlab{b}})},\ \Eprint
  {http://arxiv.org/abs/hep-th/0611259} {arXiv:hep-th/0611259 [hep-th]}
  \BibitemShut {NoStop}%
%%CITATION = HEP-TH/0611259;%%
\bibitem [{\citenamefont {Battefeld}(2008)}]{Battefeld:2007st}%
  \BibitemOpen
  \bibfield  {author} {\bibinfo {author} {\bibfnamefont {T.}~\bibnamefont
  {Battefeld}},\ }\href {\doibase 10.1103/PhysRevD.77.063503} {\bibfield
  {journal} {\bibinfo  {journal} {Phys.Rev.}\ }\textbf {\bibinfo {volume}
  {D77}},\ \bibinfo {pages} {063503} (\bibinfo {year} {2008})},\ \Eprint
  {http://arxiv.org/abs/0710.2540} {arXiv:0710.2540 [hep-th]} \BibitemShut
  {NoStop}%
%%CITATION = ARXIV:0710.2540;%%
\bibitem [{\citenamefont {Lehners}\ and\ \citenamefont
  {Renaux-Petel}(2009)}]{Lehners:2009ja}%
  \BibitemOpen
  \bibfield  {author} {\bibinfo {author} {\bibfnamefont {J.-L.}\ \bibnamefont
  {Lehners}}\ and\ \bibinfo {author} {\bibfnamefont {S.}~\bibnamefont
  {Renaux-Petel}},\ }\href {\doibase 10.1103/PhysRevD.80.063503} {\bibfield
  {journal} {\bibinfo  {journal} {Phys.Rev.}\ }\textbf {\bibinfo {volume}
  {D80}},\ \bibinfo {pages} {063503} (\bibinfo {year} {2009})},\ \Eprint
  {http://arxiv.org/abs/0906.0530} {arXiv:0906.0530 [hep-th]} \BibitemShut
  {NoStop}%
%%CITATION = ARXIV:0906.0530;%%
\end{thebibliography}%

\end{document}